# Lacunar fractal photon sieves

F. Giménez[1], W.D. Furlan[2], J.A. Monsoriu[3*]

[1]*Departamento de Matemática Aplicada, Universidad Politécnica de Valencia, E-46022 Valencia, Spain*
[2]*Departamento de Óptica, Universidad de Valencia, E-46100 Burjasstot (Valencia), Spain*
[3]*Departamento de Física Aplicada, Universidad Politécnica de Valencia, E-46022 Valencia, Spain*

**Abstract**

We present a new family of diffractive lenses whose structure is based on the combination of two concepts: photon sieve and fractal zone plates with variable lacunarity. The focusing properties of different members of this family are examined. It is shown that the sieves provide a smoothing effect on the higher order foci of a conventional lacunar fractal zone plate. However, the characteristic self-similar axial response of the fractal zone plates is always preserved.

## 1. Introduction

Diffractive focusing elements are most beneficial in many applications where they can perform tasks that are difficult, or even impossible, with conventional refractive optics [1]. In fact, during the last years, they are becoming key elements in obtaining images in several scientific and technological areas such as, THz tomography, astronomy and soft X-ray microscopy [2-5]. Aside from conventional Fresnel zone plates, other geometries of 2D photonic-image-forming structures have been proposed. One of them is generally known as fractal zone plate (FZP) [6, 7]. A FZP is characterized by its fractal profile along the square of the radial coordinate. When illuminated by a parallel wavefront a FZP produce multiple foci, the main lobe of which coincide with those of the associated conventional zone plate (actually, under certain circumstances, a FZP can be thought as a conventional zone plate with certain missing zones), but he internal structure of each focus exhibits a characteristic fractal structure, reproducing the self-similarity of the originating FZP. It has been shown that the number of foci and their relative amplitude can be modified with the lacunarity of the FZP [7]. This design parameter is frequently used as a measure of the "texture" of fractal structures [8-11].

Photon sieves [12] are another kind of diffractive optical elements, developed for focusing and imaging soft X rays with high resolution capabilities. A conventional photon sieve is essentially a Fresnel zone plate where the clear zones are replaced by great number of non overlapping holes of different sizes. Several theoretical and experimental works related to photon sieves have been performed, explaining their properties from different points of view [13, 14] and showing their good performance in different applications [15, 16]. Following this trend, our group has recently introduced the fractal photon sieve (FPS) [17] which essentially is a FZP in which the clear zones have been replaced by a non overlapping distribution of circular holes.

---

[*] *E-mail address*: jmonsori@fis.upv.es

The original proposal for FPS was based on the triadic regular Cantor fractal set [17]. In this paper we propose a more general structure introducing the lacunarity as a new design parameter. We coined this new family of diffractive structures: Lacunar Fractal Photon Sieves (LFPS). Some practical considerations abut the design of this general type of photon sieves are investigated, taking into account the physical limits imposed by the different construction parameters. The impact on the axial irradiance produced by the combination of sieves and lacunar FZP is compared with the response of the equivalent continuous lacunar FZP.

## 2. Lacunar fractal photon sieve design

The construction of a typical *polyadic Cantor fractal set* with a specific lacunarity is shown in Fig. 1. The first step in the construction procedure consists in defining a straight-line segment of unit length called *initiator* (stage $S=0$). Next, at stage $S=1$, the *generator* of the set is constructed by $N$ ($N=4$ in the figure) non-overlapping copies of the initiator each one with a scale $\gamma<1$. At the following stages of construction of the set ($S=2,3,...$), the generation process is repeated over and over again for each segment in the previous stage.

To characterize the resulting Cantor set, as well as many other fractal structures, one of the most frequently-used descriptors is the *fractal dimension,* defined as

$$D = -\ln(N)/\ln(\gamma). \tag{1}$$

However, this parameter does not uniquely define the fractal. In fact, for the general case, it is necessary to introduce another parameter to specify the distribution of the $N$ copies into the unit length segment. This parameter specifies the *lacunarity* of the resulting structure and it is necessary to complete the characterization of the fractal since structures with different lacunarity can have the same fractal dimension. To define the lacunarity, we use the width of outermost gap in the first stage (see Fig. 1). This convention was also adopted in some previous papers dealing with Cantor fractals [7, 8, 9]. It has been shown that regular fractal is a particular case of this general structure when the lacunarity $\varepsilon$, takes the following value

$$\varepsilon_R = \frac{1-N\gamma}{N-1}, \tag{2}$$

this is equivalent to impose that the bars and gaps have the same size at the initiator stage.

Assuming that the fractal structure of Fig. 1 can be mathematically represented by a one dimensional binary function $q(\varsigma)$ defined in the interval $[0,1]$, lacunar FZPs can be generated by performing a change of coordinates $\varsigma=(r/a)^2$ and by rotating the transformed 1-D function around one of its extremes. The result is a zone plate having a radial coordinate $r$ and an outermost ring of radius $a$ [see Fig. 2(a)]. The LFPS here proposed has essentially the same structure of a lacunar FZP but instead of transparent rings the corresponding zones have been broken up into isolated circular holes randomly distributed (a photon sieve). The result is shown in Fig. 2(b).

In the construction procedure we adopted the results reported in Ref [12] where it has been shown that for a photon sieve constructed with a Fresnel zone plate structure, the diameter $d$ of the holes in each ring of width $w$ of has an optimum value for the effective contribution to the focus. This value is given by $d=1.53w$. The number of holes in the LFPS shown in Fig. 2(b) is 2357 and the density of holes per zone, i.e., the ratio between the area covered by the holes and the total area of the zone, is approximately 90%. The minimum diameter, $d$, in the outermost ring is given by $d/a=0.0078$.

## 3. Axial Irradiance provided by LFPS

To investigate the axial focusing properties of a typical LFPS let us start by considering the irradiance along the optical axis, $z$, provided by an optical system having a 2D pupil function $p(r,\phi)$ (expressed in canonical polar coordinates) when it is illuminated by a plane wave of wavelength $\lambda$:

$$I(z) = \left(\frac{2\pi}{\lambda z}\right)^2 \left| \int_0^a \int_0^{2\pi} p(r,\phi) \exp\left(-i\frac{\pi}{\lambda z}r^2\right) r \, dr \, d\phi \right|^2 \tag{3}$$

The axial irradiance in Eq. (3) can be conveniently expressed in terms of a single radial integral, by performing first the azimuthal average of the pupil function $p(r,\phi)$:

$$p_o(r) = \frac{1}{2\pi} \int_0^a p(r,\phi) \, d\phi. \tag{4}$$

Then Eq. (3) can be rewritten as

$$I(z) = \left(\frac{2\pi}{\lambda z}\right)^2 \left| \int_0^a p_o(r) \exp\left(-i\frac{\pi}{\lambda z}r^2\right) r \, dr \right|^2. \tag{5}$$

To compare the performance of a LFPS with the associated lacunar FZP we used Eq. (5) to compute the axial irradiances provided by the pupil functions represented in Figs. 2(a) and 2(b). The result is shown in Fig. 3 together with the irradiances corresponding to analogous pupils constructed for $S=1$. In order to observe the fractal behavior of the irradiances for different stages of growth $S$, we have normalized the axial distance, $z$, to the principal focal length given by

$$f_S = \frac{a^2}{\lambda(2N-1)^S}. \tag{6}$$

The self-similarity of the resulting axial irradiances can be clearly seen in Fig. 3(a). The dotted lines ($S=1$) forms the envelope of the solid lines ($S=2$). On the other hand, when the lacunar FZP is replaced by a LFPS the principal focus remains almost unaltered but all odd higher orders are highly reduced due to the smoothing effect that the holes produce on the azimuthal average of the effective pupil. This effect is obtained at the expense of the appearance of low intensity even orders, which by design are null in the case of azimuthally uniform FZP with $\varepsilon=\varepsilon_R$.

To show that the self-similar behavior provided by LFPS is retained even when the lacunarity parameter is varied, we have used *twist plots* [9] may be used. These plots represent the axial irradiance is represented as a function of the normalized axial distance and the lacunarity parameter $\varepsilon$. Figure 4 shows twist plots for tetraedic ($N=4$) LFPS $S=1$ (a) and $S=2$ (b). In these plots a linear gray level scale was used for the normalized axial irradiance.

The most noticeable feature of Fig. 4 is the self-similarity that can be observed between the plots corresponding to $S=1$ and $S=2$. In effect, the rescaled data at stage $S=1$ forms an envelope for the data at $S=2$, and both structures are self-similar for any value of $\varepsilon$. This result shows that the axial irradiance provided by polyadic LFPS has self-similar properties like to those reported for lacunar FZP [7]. However, from the focalization point of view LFPS are more efficient than lacunar FZP because the relative intensity of the main focus is much higher than the secondary foci. To show more clearly this feature we represented the data in Fig. 4(b) for $S=2$ but in a 3D plot shown in Fig. 5 (bottom). In the upper part of this figure, the irradiance provided by an equivalent lacunar FZP is shown for comparison. The reduction of the secondary foci is significant for any value of the parameter $\varepsilon$, however the structure of the principal focus remain almost unaltered.

## 4. Conclusions

Fractal photon sieves with variable lacunarity have been proposed. The focusing properties of LFPSs have been numerically evaluated. As this new kind of zone plates are based on lacunar FZP, they have a principal focus with almost the same structure. However, since the azimuthal average governs the behaviour of the axial irradiance, a LFPS can be designed to achieve any convenient apodization by a proper modulation of the density of pinholes in each zone. It was shown that the lacunarity has a dramatic effect on the axial irradiance provided the LFPS, although the apodization of the higher order foci and the self-similarity of the principal focus are always preserved.

Since from a structural point of view, the LFPSs can be constructed without any supporting substrate, they can be also employed as a versatile focusing device in other regions of the electromagnetic spectrum, such as microwaves and X rays and even with slow neutrons, in which graded amplitude pupils are difficult or even impossible to construct.

**Acknowledgments**

This work was funded by the Plan Nacional I+D+i (grant DPI 2006-8309), Ministerio de Educación y Ciencia, Spain. We also acknowledge the financial support from the Programa de Incentivo a la Investigación de la UPV 2005, Vicerrectorado de Innovación y Desarrollo, Universidad Politécnica de Valencia, Spain.

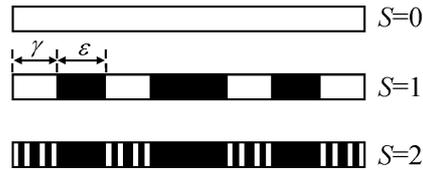

**Figure 1.** Schemes for the generation of a polyadic Cantor fractal set for $N=4$ up to $S=2$. $\gamma$ is the scale factor and $\varepsilon$ is the parameter that characterizes the lacunarity.

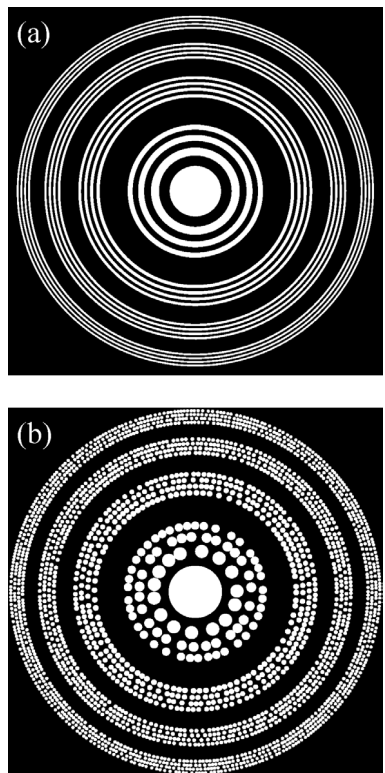

**Figure 2.** Comparison between (a) FZP and (b) FPS generated with the following parameters: $S=2$, $N=4$, $\gamma=1/7$, and $\varepsilon=\varepsilon_R=1/7$.

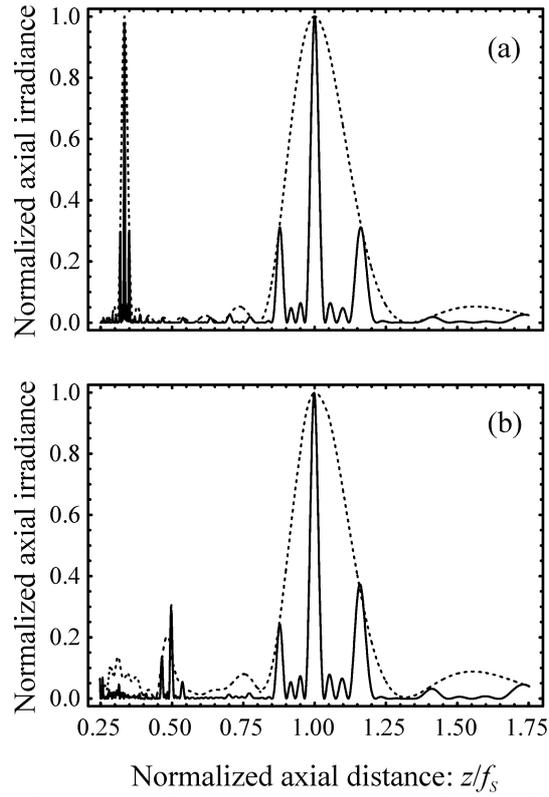

**Figure 3.** Normalized irradiance distribution along the optical axis produced by (a) FZP and (b) FPS that are shown in Fig. 2(a) and Fig. 2(b), respectively. The dotted lines represent the corresponding axial irradiances for $S=1$.

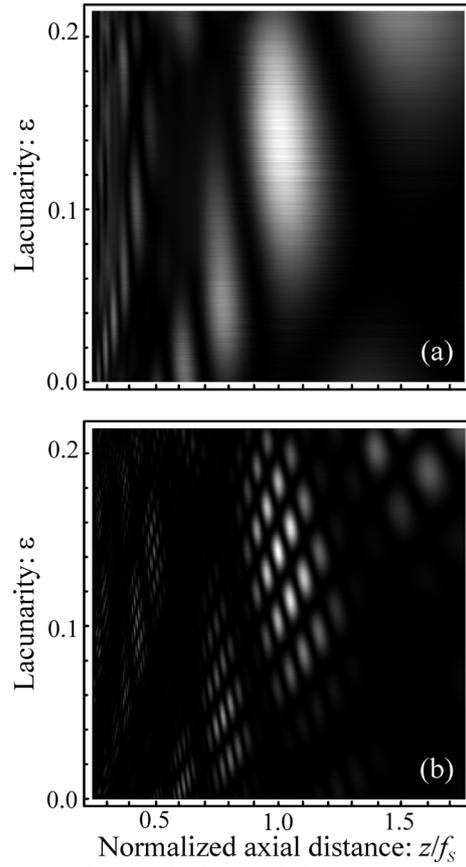

**Figure 4.** Gray-scale representation of the axial response plotted as a function of the normalized axial distance and the lacunarity of a LFPS ($N=4$, $\gamma=1/7$) for: (a) $S=1$, and (b) $S=2$.

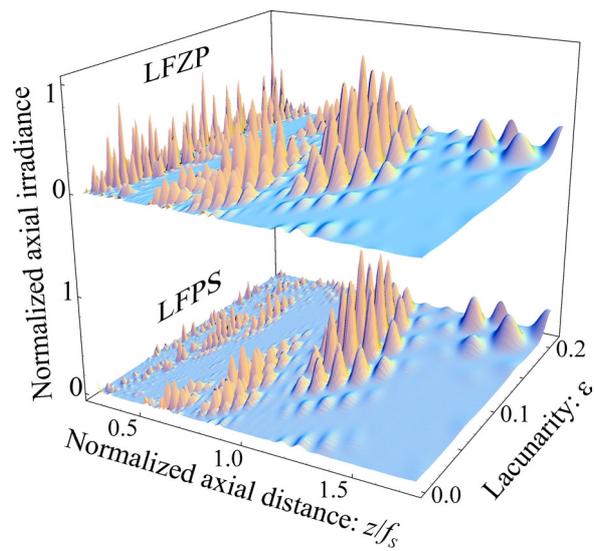

**Figure 5.** Three-dimensional representation of the the axial response plotted as a function of the normalized axial distance and the lacunarity for FZP (top) and LFPS (bottom).

7